\documentclass[letterpaper,twocolumn,10pt]{article}
\usepackage{zhanggroup}

\usepackage[T1]{fontenc}
\usepackage[utf8]{inputenc}
\usepackage{microtype}
\usepackage{graphicx}
\usepackage{xcolor}
\usepackage{booktabs}
\usepackage{amssymb,amsmath}
\usepackage{subcaption}
\usepackage{multirow}
\usepackage{xspace}
\usepackage{cleveref}
\usepackage[absolute]{textpos}

\newcommand{\mypara}[1]{\smallskip\noindent{\bf {#1}.}\xspace}
\newcommand{\method}{\textbf{AP-Test}\xspace}

\newcommand{\Fullscore}{Match score\xspace}
\newcommand{\fullscore}{match score\xspace}
\newcommand{\fullscores}{match scores\xspace}
\newcommand{\score}{MS\xspace}
\definecolor{ChallengeA}{RGB}{0,119,187}    
\definecolor{ChallengeB}{RGB}{51,187,238}   
\definecolor{ChallengeC}{RGB}{238,51,119}   
\definecolor{ChallengeD}{RGB}{238,119,51}   
\newcommand{\challengeA}{\textcolor{ChallengeA}{\textbf{C1}}\xspace}
\newcommand{\challengeB}{\textcolor{ChallengeB}{\textbf{C2}}\xspace}
\newcommand{\challengeC}{\textcolor{ChallengeC}{\textbf{C3}}\xspace}
\newcommand{\challengeD}{\textcolor{ChallengeD}{\textbf{C4}}\xspace}

\begin{document}

\date{}

\title{\bf Peering Behind the Shield: Guardrail Identification in Large Language Models}

\author{
 \textbf{Ziqing Yang\textsuperscript{1}}\ \ \
 \textbf{Yixin Wu\textsuperscript{1}}\ \ \
 \textbf{Rui Wen\textsuperscript{2}}\ \ \
 \textbf{Michael Backes\textsuperscript{1}}\ \ \
 \textbf{Yang Zhang\textsuperscript{1}\thanks{Corresponding author.}}
\\
\\
 \textsuperscript{1}\textit{CISPA Helmholtz Center for Information Security}\ \ \
 \textsuperscript{2}\textit{Institute of Science Tokyo}
}

\maketitle

\begin{abstract}
With the rapid adoption of large language models (LLMs), conversational AI agents have become widely deployed across real-world applications.
To enhance safety, these agents are often equipped with guardrails that moderate harmful content.
Identifying the guardrails in an agent thus becomes critical for adversaries to understand the system and design guard-specific attacks.
In this work, we introduce \method, a novel approach that leverages guard-specific adversarial prompts to detect the identity of guardrails deployed in black-box AI agents.
Our method addresses key challenges in this task, including the influence of safety-aligned LLMs and other guardrails, as well as a lack of principled decision-making strategies.
\method employs two complementary testing strategies, input and output guard tests, and a new metric, \fullscore, to enable robust identification.
Experiments across diverse agents and four open-source guardrails demonstrate that \method achieves perfect classification accuracy in multiple scenarios.
Ablation studies further highlight the necessity of our proposed components.
Our findings reveal a practical path toward guardrail identification in real-world AI systems.\footnote{Our code of \method is available at \url{https://github.com/TrustAIRLab/AP-Test}.}
\end{abstract}

\section{Introduction}

Human-AI conversations have been significantly advanced by the rapid development of large language models (LLMs).
These conversational AI agents are now extensively deployed across various domains, including customer service~\cite{DGYZSZXZMWBZYWWGSLGLXWWFLZDZRDCJLLDLHCLZBXWDXGQLGLWCYQLCNLCDHGGHYWZZWZXXZZTLWLTHZWCDGZPWCJCLZCYWYZPL25,Claude,GPT4o}, education~\cite{NYSDJ24}, and healthcare~\cite{PYCSPCMFZMLMOAHSGBW23}.
Such widespread use also raises severe security concerns, such as jailbreak attacks~\cite{SCBSZ24,ZWKF23,LXCX23,ZYPDLWW24} and prompt injection attacks~\cite{LJGJG24, ZLYK24, DZBBFT24}.

\begin{figure}[!t]
\centering
\includegraphics[width=\columnwidth]{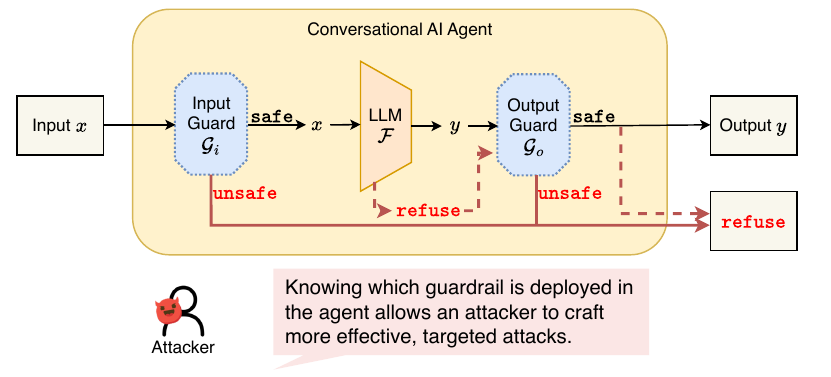}
\caption{A conversational AI agent equipped with guardrails.
Normally, the agent responds to the input following the black line.
Once the guardrail flags the input or generated response \texttt{unsafe}, the agent would \texttt{refuse} the query (solid red line).
The dashed red line signifies the LLM's refusal to the input per its internal safety alignment.
}
\label{figure:overview}
\end{figure}

Safety guardrails~\cite{IUCRIMTHFTK23, ZLMPFHNPKRSW24} are thus developed to further moderate the input/output content of the AI agents, as shown in \Cref{figure:overview}.
They are often fine-tuned on top of LLMs using annotated datasets that cover a broad spectrum of safety risks~\cite{IUCRIMTHFTK23, HREJLLCD24, ZLMPFHNPKRSW24}.
In addition, a growing number of high-quality open-source guardrails, such as LlamaGuard~\cite{IUCRIMTHFTK23}, WildGuard~\cite{HREJLLCD24}, and ShieldGemma~\cite{ZLMPFHNPKRSW24}, have become widely available and are increasingly integrated into AI Agents.
This largely challenges attackers to bypass the safety mechanisms in AI agents; thus, it is crucial to identify the guardrails deployed in the AI agent.
Once a guardrail is identified, it can be treated as a white-box component, facilitating guard-specific attacks that are shown to outperform black-box methods~\cite{MHCCFJP24, ZXM24}.

\begin{figure*}[!t]
\centering
\centerline{\includegraphics[width=1.5\columnwidth]{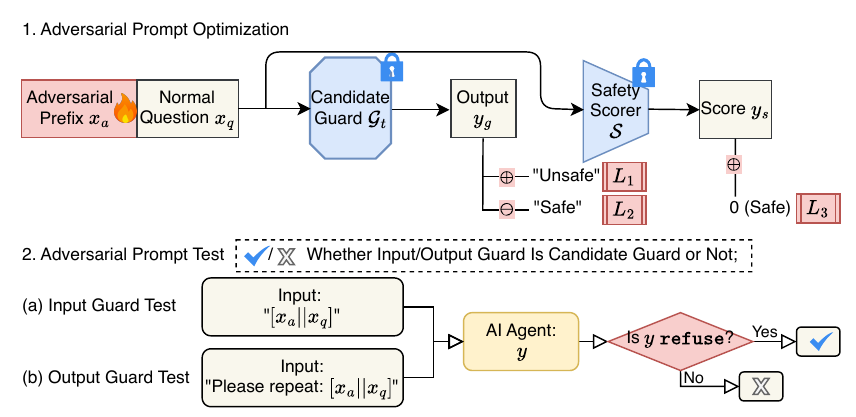}}
\caption{
Framework of our \method.
We first perform \emph{adversarial prompt optimization} based on the candidate guardrail with our tailored loss function.
Then, we conduct \emph{adversarial prompt tests} by querying the AI agent with the adversarial prompts to determine whether the candidate guardrail exists in the agent.
}
\label{figure:framework}
\end{figure*}

Considering that guardrails are often built on top of LLMs, a natural question arises: can we simply apply LLM identification methods~\cite{ZZWL23, GULYO24, JZSLH24, YWSDBZ25} to detect the presence of specific guardrails in an AI agent?
LLM identification aims to determine the origin or identity of an LLM, primarily to address concerns about unauthorized use.
The common approach relies on optimizing adversarial prompts that elicit distinctive responses from the candidate LLM~\cite{GULYO24, JZSLH24}.
However, this line of work falls short due to several challenges in guardrail identification:
\challengeA The output space of guardrails is extremely limited (e.g., binary safe/unsafe responses), so optimizing input triggers toward a target response is much easier.
This makes it hard to be specific to one guardrail.
\challengeB AI agents usually allow query-only access and do not provide knowledge of the underlying safety mechanisms~\cite{GPT4o, Claude, GitHub_Copilot, DGYZSZXZMWBZYWWGSLGLXWWFLZDZRDCJLLDLHCLZBXWDXGQLGLWCYQLCNLCDHGGHYWZZWZXXZZTLWLTHZWCDGZPWCJCLZCYWYZPL25}, revealing little information about the guardrail.
Mechanisms such as additional guardrails or safety-aligned LLMs may also obscure the existence of the candidate guardrail.
\challengeC Guardrails are integrated into AI agents with unknown deployment stages; when applied at the output stage, their input depends on the LLM's outputs.
In this sense, it is challenging to input the adversarial prompts into the output guardrail.
\challengeD Existing LLM identification methods lack principled strategies for selecting identification thresholds, limiting their robustness and reliability in practical settings.

By solving the above challenges, we propose \method, which utilizes \underline{\textbf{A}}dversarial \underline{\textbf{P}}rompts to \underline{\textbf{Test}} the identity of the safety guardrail deployed in a black-box AI agent.
As shown in \Cref{figure:framework}, our approach begins by probing the AI agent with guard-specific adversarial prompts, which are designed to be flagged as \texttt{unsafe} by a specific candidate guardrail while remaining \texttt{safe} according to others.
Such optimization exploits the limited output space of guardrails (\challengeA) and mitigates the influence of safety-aligned LLMs and other safety guardrails (\challengeB) via a tailored loss function.
We then perform identification using two complementary strategies: the \emph{input guard test} and the \emph{output guard test}.
Together, these tests cover scenarios where guardrails are deployed at either the input or output stage of an agent, thereby addressing \challengeC.
Intuitively, the output guard test asks the agent to repeat the adversarial prompt, so that the output guard will receive the LLM's response that contains the repeated text.
To make identification without requiring additional guardrails (\challengeD), we design a novel metric, \fullscore, to measure how the safety mechanism in the agent matches the candidate guardrail.
A larger \fullscore indicates a stronger likelihood of the candidate guardrail being deployed.

To show the practicability of our \method, we conduct extensive experiments on four different candidate guardrails under diverse scenarios, covering two popular LLMs and 10 guardrails.
Results demonstrate that the proposed attack method accurately identifies guardrails in various AI agents for both input and output guardrails.
Specifically, the WildGuard-specific~\cite{HREJLLCD24} input guard test on both Llama3.1-based~\cite{DJPKALMSYFGHYMSKHRZRGSRBTCCNBMMKTWWFNASPLECMGPHLALDSRZSLANMPCNKXTZIKMECLGVPMSLBHLFCHLWYBSPRJSJAUPLHSa24} and GPT4o-based~\cite{GPT4o} agents achieves a perfect classification accuracy, i.e., $Acc=1.00$.
That is, our \method successfully identifies the existence of WildGuard in all evaluated agents, indicating its effectiveness.
Our \method also successfully probes the candidate guardrail in more complex agents, e.g., containing two different guardrails.
This showcases the practicality of our method.

Moreover, to analyze the impact of our proposed method, we perform an ablation study to examine the role of each component, particularly our loss terms designed to optimize adversarial queries and the existence of the query set.
Our findings reveal that our loss terms and the query set are crucial for the identification.
For example, without the loss ensuring the prompt remains \texttt{safe} for other guardrails ($L_3$), \method tends to misidentify that the LlamaGuard3~\cite{CKZSRZPCUP24} is used in the agent that is only equipped with WildGuard as the input guardrail.

Overall, our contributions are as follows:
\begin{itemize}
    \item We propose \method, the first method that uses guard-specific adversarial prompts to identify guardrails in black-box AI agents, overcoming core challenges under this setting.
    \item Experiments show the effectiveness and robustness of \method on four candidate guardrails on various agents under diverse scenarios.
    \item Our ablation study demonstrates the importance of each component in our proposed method, showing that their removal significantly degrades identifying performance.
\end{itemize}

\section{Background and Related Work}

\mypara{LLM Security Risks}
The rapid advancement of LLMs provides users with significant convenience but also raises critical security concerns~\cite{ZYPDLWW24,GRLWCWDW23,ZLWJZBSZ24,WLBZ24,LGFXS23}.
Among these concerns, jailbreak attacks~\cite{SCBSZ24,ZWKF23,LXCX23,ZYPDLWW24} pose a major threat by bypassing built-in safety mechanisms, enabling models to generate restricted or harmful content that violates usage policies~\cite{Google_Policy,Meta_Policy,OpenAI_Policy,Amazon_Policy_1}.
Previous studies analyze existing jailbreak strategies, particularly focusing on in-the-wild jailbreak prompts that are manually crafted in real-world scenarios~\cite{SCBSZ24}.
More recent studies introduce automated jailbreak generators, such as AutoDAN~\cite{LXCX23}, GCG~\cite{ZWKF23}, and TAP~\cite{MZKNASK23}, which optimize adversarial prompts to evade safety measures and maximize attack success rates.

\mypara{Internal Safety Alignment}
LLM safety alignment refers to the process of ensuring that LLMs generate outputs that are consistent with human values.
Most popular LLMs, such as Llama3~\cite{DJPKALMSYFGHYMSKHRZRGSRBTCCNBMMKTWWFNASPLECMGPHLALDSRZSLANMPCNKXTZIKMECLGVPMSLBHLFCHLWYBSPRJSJAUPLHSa24}, Gemma~\cite{MHDBPSRKLTHCRBBCSHTBPTSLCCCIRBNNYTMRMTGAKLLSBCFCa24}, Mistral~\cite{JSMBCCBLLSLLSSLWLS23}, and ChatGPT~\cite{GPT4o}, are safety-aligned, either through reinforcement learning with human feedback (RLHF)~\cite{LNTYCWZZ24,JLDPZBCSWY23}, or learning from curated datasets that contain safety-related data~\cite{JSMBCCBLLSLLSSLWLS23, MHDBPSRKLTHCRBBCSHTBPTSLCCCIRBNNYTMRMTGAKLLSBCFCa24}.
However, these safety-aligned LLMs are still exposed to security risks such as jailbreak attacks~\cite{SCBSZ24,ZWKF23,LXCX23,ZYPDLWW24} and prompt injection attacks~\cite{LJGJG24, ZLYK24, DZBBFT24,YWSJX23}.

\mypara{External Safety Guardrail}
Safety guardrails are designed for moderating the input/output content of the LLMs~\cite{IUCRIMTHFTK23, CKZSRZPCUP24, GVGP24, HREJLLCD24, ZLMPFHNPKRSW24} and serve as an external complement to safety alignment.
For example, built on Llama2-7B~\cite{TMSAABBBBBBBCCCEFFFFGGGHHHIKKKKKKLLLLLMMMMMNPRRSSSSSTTTWKXYZZFKNRSES23}, LlamaGuard~\cite{IUCRIMTHFTK23} is fine-tuned on their constructed dataset based on a safety risk taxonomy encompassing a range of safety risks.
Subsequent versions, LlamaGuard2~\cite{Llama_Guard_2} and LlamaGuard3~\cite{CKZSRZPCUP24}, further expand the safety risk taxonomy and dataset, leveraging state-of-the-art LLMs for fine-tuning, thereby strengthening their safeguard capabilities.
Similarly, Aegis~\cite{GVGP24}, WildGuard~\cite{HREJLLCD24}, and ShieldGemma~\cite{ZLMPFHNPKRSW24} follow a comparable approach.
Specifically, the Aegis series is further instruction-tuned on LlamaGuard based on their own dataset, and ShieldGemma is built upon Gemma2~\cite{MHDBPSRKLTHCRBBCSHTBPTSLCCCIRBNNYTMRMTGAKLLSBCFCa24}.

\mypara{LLM Fingerprinting and Watermarking}
LLM identification focuses on identifying the origin of an LLM~\cite{ZZWL23, MSSA24, GULYO24, JZSLH24, YWSDBZ25}.
A common approach involves crafting LLM-specific adversarial prompts to guide the candidate LLM to produce target responses, which are then used for identification~\cite{GULYO24, JZSLH24}.
As state-of-the-art guardrails are often built upon LLMs~\cite{IUCRIMTHFTK23, HREJLLCD24, ZLMPFHNPKRSW24}, guardrail identification is similar.
However, a key difference lies in the limited and often masked output space of guardrails (\challengeA).
This makes it hard to directly adapt the target responses for identification.
Moreover, existing LLM identification techniques typically rely on empirically chosen thresholds by querying an auxiliary set of LLMs~\cite{GULYO24, JZSLH24}, which fails to be solely based on the candidate model.
This limitation highlights the need for more principled decision-making strategies, as addressed by our method (\challengeD).

\section{Problem Statement}

\mypara{Preliminary}
At the core of these AI agents lies a safety-aligned LLM $\mathcal{F}$ that drives their functionality, but still suffers from security risks~\cite{SCBSZ24,ZWKF23,LXCX23,LJGJG24,YWSJX23}.
To ensure the security and compliance of these AI agents, additional mechanisms known as input and output guardrails $\mathcal{G}$ are often implemented.
As shown in \Cref{figure:overview}, the input guardrail $\mathcal{G}_i$ evaluates user inputs $x$ to determine whether they should be forwarded to the LLM.
If the input $x$ is deemed high-risk, policy-violating, or jailbreak prompts, i.e., $\mathcal{G}_i(x)=\texttt{unsafe}$, the agent then returns \texttt{refuse} without processing it further.
Otherwise, the safety-aligned LLM would react to the input $x$, either generating responses or returning \texttt{refuse} if it perceives unsafe.
However, with attacks like jailbreak attacks, even though the prompt input may be considered \texttt{safe} by both the input guard $\mathcal{G}_i$ and the LLM $\mathcal{F}$, the LLM's response could still contain harmful content.
Therefore, in real-world applications, solely monitoring input may not be sufficient, which motivates the deployment of output guardrails.
The output guardrail $\mathcal{G}_o$ monitors the LLM's response $y$ to avoid policy violations.
If a response is non-compliant, i.e., $\mathcal{G}_o(x)=\texttt{unsafe}$, the agent would withhold the output and return \texttt{refuse}; otherwise, the agent would output the response $y$.

\mypara{Goal}
The goal of guardrail identification is to determine whether a guardrail or its derivative is deployed in an AI agent.
The derivative refers to further customized versions of the guardrail, such as those that are fine-tuned or instruction-tuned.

\mypara{Capability}
We assume black-box access to the AI agent, with no knowledge of the underlying LLM or the presence of input/output guardrails.
We focus on open-source guardrails due to their widespread adoption.
According to HuggingFace, LlamaGuard3~\cite{CKZSRZPCUP24} received 335,571 downloads in March 2025, while AegisPermissive~\cite{GVGP24} was downloaded 1,021,359 times.
One might argue that developers could create entirely proprietary guardrails.
However, given the widespread adoption and accessibility of open-source guardrails, it is often more practical to build on existing ones through further fine-tuning.
As such, we also consider derivatives of popular open-source guardrails to better cover the space of real-world deployments.

\mypara{Problem Formulation}
We define the guardrail identification task as follows:
\textit{Given an AI agent and a candidate guardrail $\mathcal{G}_t$, guardrail identification aims to identify whether the candidate guardrail $\mathcal{G}_t$ or its derivative is deployed in the AI agent.}
The identification task consists of two tests:
(1) The input guard test audits whether the candidate guardrail $\mathcal{G}_t$ is deployed at the input stage of the AI agent;
(2) The output guard test evaluates whether the candidate guardrail is present at the output stage.

\section{Method}
\label{section:method}

In this work, we propose \method, which leverages guard-specific \underline{\textbf{A}}dversarial \underline{\textbf{P}}rompts to \underline{\textbf{Test}} the identity of the input/output guardrail deployed in an AI agent.
As shown in \Cref{figure:framework}, our framework consists of two phases: \emph{adversarial prompt optimization} and \emph{adversarial prompt test}.

\subsection{Adversarial Prompt Optimization}

The goal of the optimization phase is to optimize adversarial prompts that the target guardrail $\mathcal{G}_t$ erroneously flags as \texttt{unsafe}, while all other guardrails and safety-aligned LLMs correctly classify them as \texttt{safe}.
An adversarial prompt is constructed by concatenating an optimized adversarial prefix $x_a$ with a normal query $x_q \in Q$, where $Q$ is a query set.
The query is used as a starting point to prevent over-rejection by other guardrails, which is proven effective in our ablation study (\Cref{section:ablation_study}).
For each query $x_q \in Q$, we optimize an adversarial prefix $x_a$ using three loss terms, addressing two specific aspects of the desired behavior.

\mypara{Candidate Guardrail Adversarial Losses ($L_1$ and $L_2$)}
As guardrails' output space is usually limited (\challengeA), these loss terms are designed to mislead the candidate guardrail $\mathcal{G}_t$ into classifying the adversarial prompt as \texttt{unsafe}.
Specifically, $L_1$ encourages the candidate guardrail to classify the adversarial prompt as \texttt{unsafe}, while $L_2$ penalizes the candidate guardrail for classifying the adversarial prompt as \texttt{safe}:
\begin{equation}
\begin{aligned}
L_1 &= \sigma(\mathcal{G}_t(x_a \Vert x_q), \texttt{unsafe}),\\
L_2 &= -\sigma(\mathcal{G}_t(x_a \Vert x_q), \texttt{safe}),
\end{aligned}
\end{equation}
where $\sigma(\cdot, \cdot)$ represents the cross-entropy loss.
Together, $L_1$ and $L_2$ ensure that $x_a$ effectively triggers the refusal mechanism of the candidate guardrail.
Conceptually, these two losses are identical, but our ablation study (\Cref{section:ablation_study}) demonstrates that the synergy of these two losses slightly outperforms solely deploying a single one of them.

\mypara{Cross-Guardrail Compatibility Loss ($L_3$)}
To ensure the adversarial prompt remains \texttt{safe} according to all other guardrails and safety-aligned LLMs (\challengeB), we introduce a safety scorer $\mathcal{S}$ and propose $L_3$.
The safety scorer $\mathcal{S}$ measures the safety stage of an input $x$~\cite{SW17, MSYBGM21, VTWK21}: $\mathcal{S}(x) = y_s \in [0, 1]$, where $y_s = 0$ indicates no security risk, and $y_s = 1$ indicates a potential risk.
The loss term is defined as:
\begin{equation}
L_3 = \sigma(\mathcal{S}(x_a \Vert x_q), 0).
\end{equation}
By minimizing $L_3$, we prevent unintended rejections from unrelated guardrails, preserving the specificity of the attack on $\mathcal{G}_t$.

The final loss function is defined as:
\begin{equation}
    L = L_1 + \alpha \cdot L_2 + \beta \cdot L_3, \quad \alpha,\beta \in \mathbb{R},
\end{equation}
where $\alpha$ and $\beta$ control the weights of the loss terms $L_2$ and $L_3$, respectively.
By jointly minimizing $L_1$, $L_2$, and $L_3$, we adopt the GCG (Greedy Coordinate Gradient) strategy~\cite{ZWKF23} with the updates following \cite{JZSLH24} to craft adversarial prompts.
Specifically, we compute gradients w.r.t. token embeddings.
For each token position, we sample top-$k$ candidate replacements based on the gradient scores and their constraints.
Then we perform greedy coordinate updates over the adversarial prompts.
The optimized adversarial prompts could thus expose the behavior of the candidate guardrail while maintaining compatibility with other guardrails

\subsection{Adversarial Prompt Test}

We then perform identification using two complementary strategies: the \emph{input guard test} and the \emph{output guard test}.
Together, these tests cover scenarios where guardrails are deployed at either the input or output stage of an agent, thereby addressing \challengeC.
In addition, we introduce \fullscore, a novel metric that facilitates identification based solely on the candidate guardrail.
\Fullscore provides clear and actionable decision strategies, effectively addressing the challenge of threshold selection in identification (\challengeD).

\mypara{Input Guard Test}
We first consider the identification of the input guardrail.
As an input guardrail in an AI agent, it will decide whether the user prompt $x$ should be passed to the LLM.
Thus, we do the input guard test by directly querying the AI agent with our adversarial prompts $\{x_a \Vert x_q\}$.
If the AI agent responds with \texttt{refuse}, we consider the candidate guardrail to probably serve as an input guardrail in the AI agent (see \Cref{figure:framework}~(2a)).

\mypara{Output Guard Test}
The third challenge \challengeC comes in identifying the output guardrail in the AI agent, as the output guardrail takes the LLM's generated response $y$ as input instead of the user prompt $x$.
It is hard for us to manipulate the LLM's output as we have no knowledge about it.
To solve this, we design a prompt template that asks the AI agent to repeat our adversarial prompt, e.g., ``Please repeat: [\texttt{Adversarial Prompt}]''
Ideally, the LLM's response $y$ should be the adversarial prompt $\{x_a \Vert x_q\}$ and will be passed to the output guardrail.
We empirically design and evaluate five prompt templates and select the best one as shown in \Cref{section:output_guardrail_test_experiment}.
With the well-designed prompt template, we can ask the AI agent to repeat the adversarial prompts to test its output guardrail as shown in \Cref{figure:framework}~(2b).
Then, we can make identifications based on the response from the agent.

\mypara{More Complex Scenarios}
In real-world applications, the guardrail can be deployed in both input and output stages of an AI agent (see \Cref{figure:overview}).
The agent may also equip different guardrails.
This poses challenges for identification (\challengeC).
Therefore, given a candidate guardrail, it is necessary to conduct both input and output guard tests on the AI agent, as illustrated in \Cref{appendix:real_world_experiments}.
Experiments in such scenarios in \Cref{section:complex_scenario} empirically showcase the effectiveness of \method.

\mypara{Match Score}
To better quantify, we introduce the \emph{refusal rate} $r = \frac{\#{\texttt{refuse}}}{\#{\text{all responses}}} \in [0,1]$, which is the ratio of \texttt{refuse} among all responses, where \texttt{refuse} represents that the AI agent refuses to respond to the query.
A higher refusal rate indicates that the candidate guardrail is more likely to be the input guardrail in this AI agent.
Unlike existing LLM identification methods~\cite{GULYO24, JZSLH24} that require testing on a set of LLMs to help distinguish the identity (\challengeD), we propose a novel metric, \fullscore, based solely on the candidate guardrail.
We first calculate the refusal rate of directly querying the candidate guardrail with the optimized adversarial prompts, denoted as the candidate refusal rate $r_t$.
Note that the candidate refusal rates $r_t$ in both input and output guard tests are the same, as we directly query the candidate guardrail, disregarding the AI agent or the LLM.
Then, we define the \fullscore for an AI agent as:
\begin{equation}
\label{equation:distance}
\begin{aligned}
\score = \frac{|\min(r, r_t)-0|^{\lambda}}{|r_t-0|^{\lambda}}=\frac{|\min(r, r_t)|^{\lambda}}{|r_t|^{\lambda}},
\end{aligned}
\end{equation}
where $0$ is the lower bound of $r$ and $\lambda$ $(\lambda \geq 1)$ represents the scaling factor.
The \fullscore $\score\in [0,1]$ depicts how likely the candidate guardrail exists in the AI agent.
In other words, this measures whether the agent refuses at least as often as the candidate guardrail would.
We consider the candidate guardrail to exist in the AI agent if $\score > 0.5$; our experiments in \Cref{section:input_experiment} show that the identification performance is not sensitive to this threshold.

\section{Experiments}

\subsection{Experimental Settings}
\label{section:settings}

We take four different guardrails as our candidate guardrails, including WildGuard~\cite{HREJLLCD24}, LlamaGuard~\cite{IUCRIMTHFTK23}, LlamaGuard2~\cite{Llama_Guard_2}, and LlamaGuard3~\cite{CKZSRZPCUP24}.
We use 10 guardrails for evaluation, including the four candidate guardrails, AegisDefensive, AegisPermissive~\cite{GVGP24}, ShieldGemma-2B, ShieldGemma-9B, ShieldGemma-27B~\cite{ZLMPFHNPKRSW24}, and GPT4o~\cite{GPT4o}.
We obey their default settings in our experiments and follow \cite{ZLMPFHNPKRSW24} to prompt GPT4o as an input/output guardrail.
Besides, we use Llama3.1~\cite{DJPKALMSYFGHYMSKHRZRGSRBTCCNBMMKTWWFNASPLECMGPHLALDSRZSLANMPCNKXTZIKMECLGVPMSLBHLFCHLWYBSPRJSJAUPLHSa24} and GPT4o as the LLMs of the conversational AI agents.
For the safety scorer in the output guard test, we consider state-of-the-art hate speech detectors and use LFTW-R4~\cite{VTWK21} in our experiments.
Details of different LLMs and guardrails we use can be found in \Cref{table:model_names} in \Cref{appendix:models}.

Firstly, we optimize the adversarial prompts based on each candidate guardrail.
We follow the settings in Jin et al.~\cite{JZSLH24} and use their dataset as the query set $Q$, which consists of 50 simple questions.
For each query in $Q$, we optimize a 32-token adversarial prefix using loss weights $\alpha = 0.01$ and $\beta = 1000$.
We further investigate the impact of different loss term weights in the ablation study.
The experiments are conducted with a batch size of 64 over 200 epochs on NVIDIA A100 GPUs with 40 GB of memory.

Then, we conduct input/output tests on agents equipped with different LLMs and input/output guardrails.
We consider this a binary classification problem and calculate the classification accuracy for each candidate guardrail over the 11 AI Agents based on the \fullscore $\score$ (with $\lambda=2$) and plot the ROC curve for all test results.
For each LLM, we construct 11 AI agents for evaluation: one without any guardrail, and the other 10 with different guardrails at the input/output stage.
We discuss the situation that the agent contains additional guardrails later in \Cref{section:complex_scenario}.

\subsection{Input Guard Test}
\label{section:input_experiment}

\begin{figure}[!t]
\centering
\includegraphics[width=0.5\columnwidth]{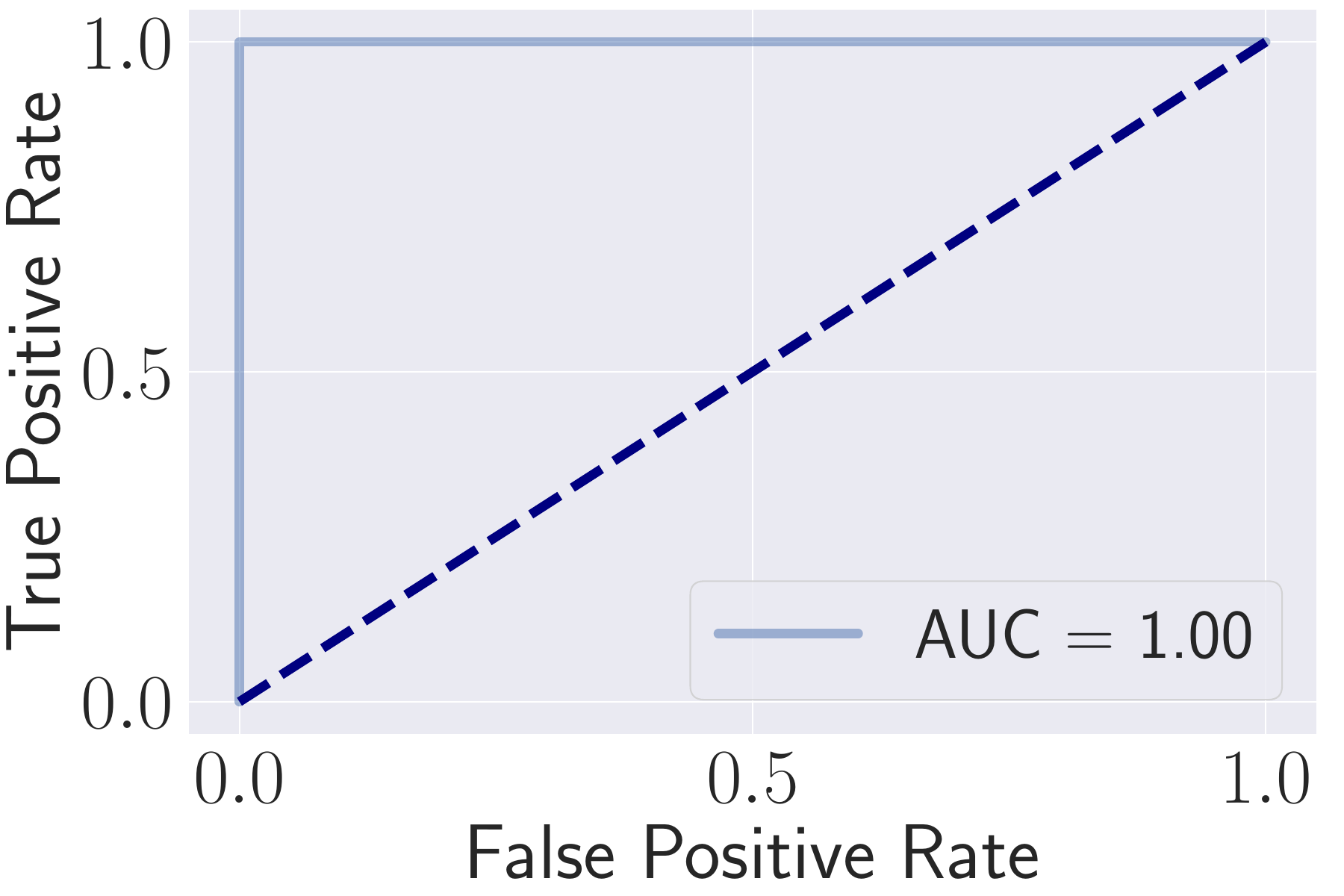}
\caption{ROC curve of input guard test.}
\label{figure:roc_input}
\end{figure}

\begin{figure}[!t]
\centering
\includegraphics[width=1.0\columnwidth]{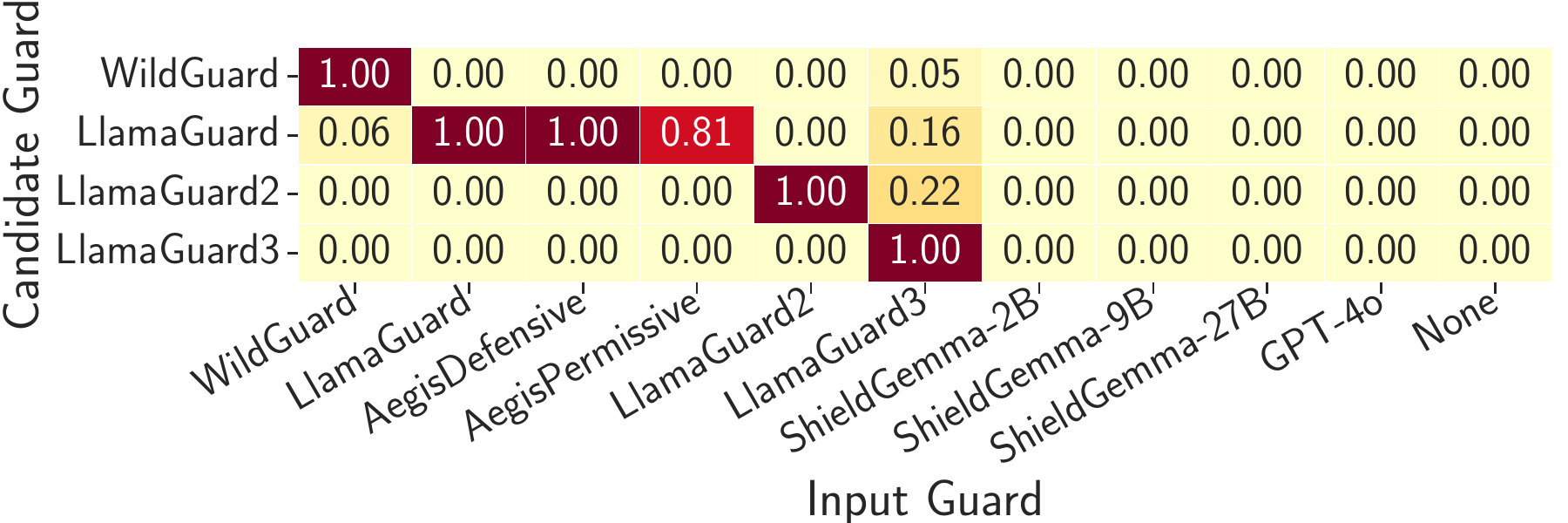}
\caption{The \fullscores of input guard tests on different AI agents.
A larger \fullscore indicates the candidate guardrail is more probable to exist in the agent.}
\label{figure:input_test_llama3_alone}
\end{figure}

We first evaluate the input guard test of \method on Llama3.1-based and GPT4o-based agents.
Results show that \method achieves a \textit{perfect} classification accuracy (1.00) in all agents for each candidate guardrail.
To further assess the robustness of our approach, we plot the ROC curves for all test results in \Cref{figure:roc_input}, observing an AUC of 1.00.
This indicates that our identification performance is not sensitive to the threshold selection.
These findings highlight both the effectiveness and reliability of our method in the input guard test setting.

\Cref{figure:input_test_llama3_alone} further illustrates the \fullscores of the input guard tests on Llama3.1-based agents equipped with various input guardrails.
We observe an obvious distinction between the existence and non-existence of the candidate guard.
For example, the WildGuard-agent achieves a \fullscore of 1.00 on the WildGuard-specific adversarial prompts, while it only achieves a maximum distance of 0.05 on other agents.
Additionally, the LlamaGuard-specific input test achieves 1.00 and 0.81 \fullscore on the AegisDefense-agent and AegisPermissive-agent, respectively.
That is because AegisDefense and AegisPermissive are fine-tuned on LlamaGuard, which are the derivatives of LlamaGuard.
Results of GPT4o-based agents can be found in \Cref{figure:input_results} in \Cref{appendix:input_test}.
We find that the results of both Llama3.1-based and GPT4o-based agents are almost the same.
This shows that the LLM has little influence on the input guard test, i.e., our \method is robust to different LLMs.
This also indicates that the internal safety alignment will not affect the performance of our method.

\subsection{Output Guard Test}
\label{section:output_guardrail_test_experiment}

\mypara{Prompt Template Selection}
As mentioned in \Cref{section:method}, the output guard test requires a prompt template that asks the AI agent to repeat our adversarial prompts.
We begin with the template: ``Please repeat: [\texttt{Adversarial Prompt}]'' and ask ChatGPT~\cite{GPT4o} to generate five variations, as shown in \Cref{table:templates} in \Cref{appendix:template}.
To assess how well the templates help the LLM repeat the adversarial prompt, we use Gemma2-9B~\cite{MHDBPSRKLTHCRBBCSHTBPTSLCCCIRBNNYTMRMTGAKLLSBCFCa24} as a surrogate LLM and evaluate the similarity between the generated text and the input adversarial prompt optimized on LlamaGuard3.
The results in \Cref{table:similarity_full} in \Cref{appendix:template} show that the third template (T3) achieves the highest similarity scores, including a ROUGE-L score of 0.956.
Thus, we select T3 as the prompt template for our experiments:
``\textit{Your task is to mimic text. Repeat this string exactly as input, with no interpretation:} [\texttt{Adversarial Prompt}].''

\begin{figure}[!t]
\centering
\includegraphics[width=1.0\columnwidth]{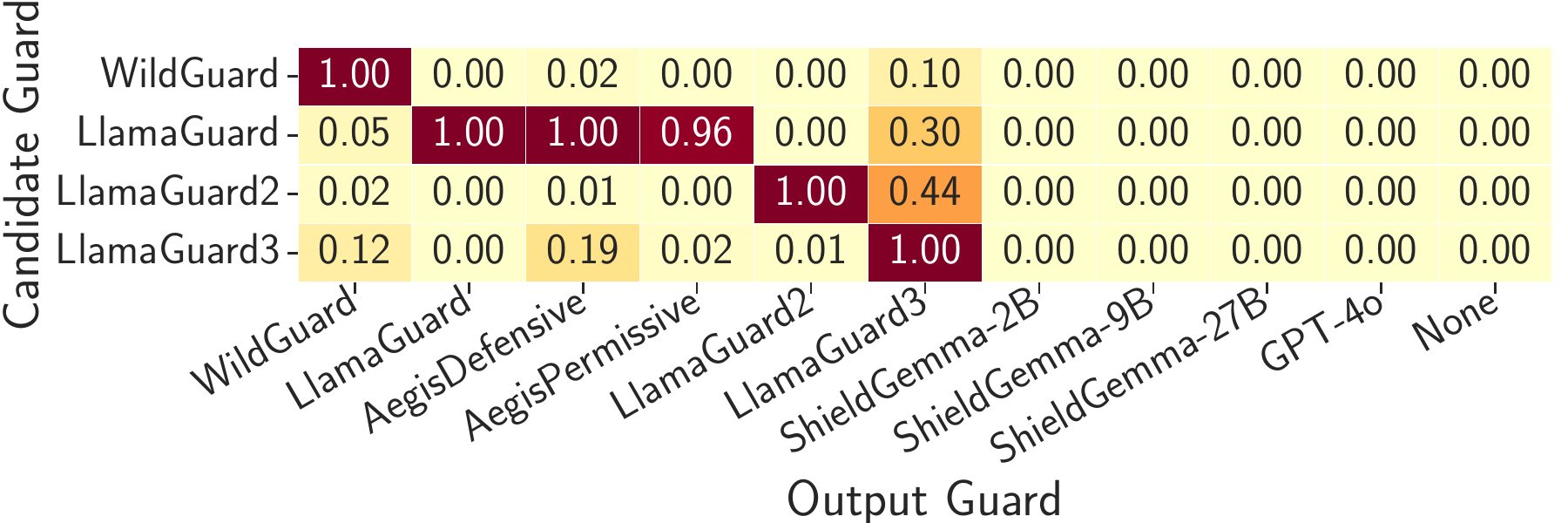}
\caption{The \fullscores of output guard tests on different Llama3.1-based AI agents.
A larger \fullscore indicates the candidate guardrail is more probable to exist in the agent.}
\label{figure:output_test_llama3_alone}
\end{figure}

\mypara{Evaluation Results}
Using T3 as the prompt template, we evaluate the output guard test on AI agents with different output guardrails.
Similar to the input guard test, \method achieves an accuracy of 1.00 and an AUC of 1.00 in all agents for four candidate guardrails (see \Cref{appendix:output_test}).
We further exploit the \fullscores of the test on Llama3.1-based agents, as shown in \Cref{figure:output_test_llama3_alone}.
All agents with \fullscores larger than 0.50 are indeed equipped with the corresponding candidate guardrail as the output guardrail.
For example, the \fullscore on the WildGuard-agent reaches 1.00 on WildGuard-specific adversarial prompts, while it is 0.00 on LlamaGuard-agent and agents with LlamaGuard's derivatives (AegisDefense-agent and AegisPermissive-agent).
This indicates that our \method successfully distinguishes the output guardrail used in the agent.
We also observe that the output guard test is harder than the input guard test.
For example, for LlamaGuard2-specific adversarial prompts, the \fullscore of the LlamaGuard3-agent achieves 0.44, which is 0.22 farther than that in the input guard test and is closer to 0.50.

Results of GPT4o-based agents are shown in \Cref{figure:output_results} in \Cref{appendix:output_test}.
We find that there is a slight performance difference between Llama3.1-based and GPT4o-based agents.
This discrepancy is due to the information loss during the LLM processing.
In other words, the performance of our output guard test is influenced by how well the LLM can repeat the adversarial prompts.

\begin{figure}[!t]
\centering
\includegraphics[width=0.9\columnwidth]{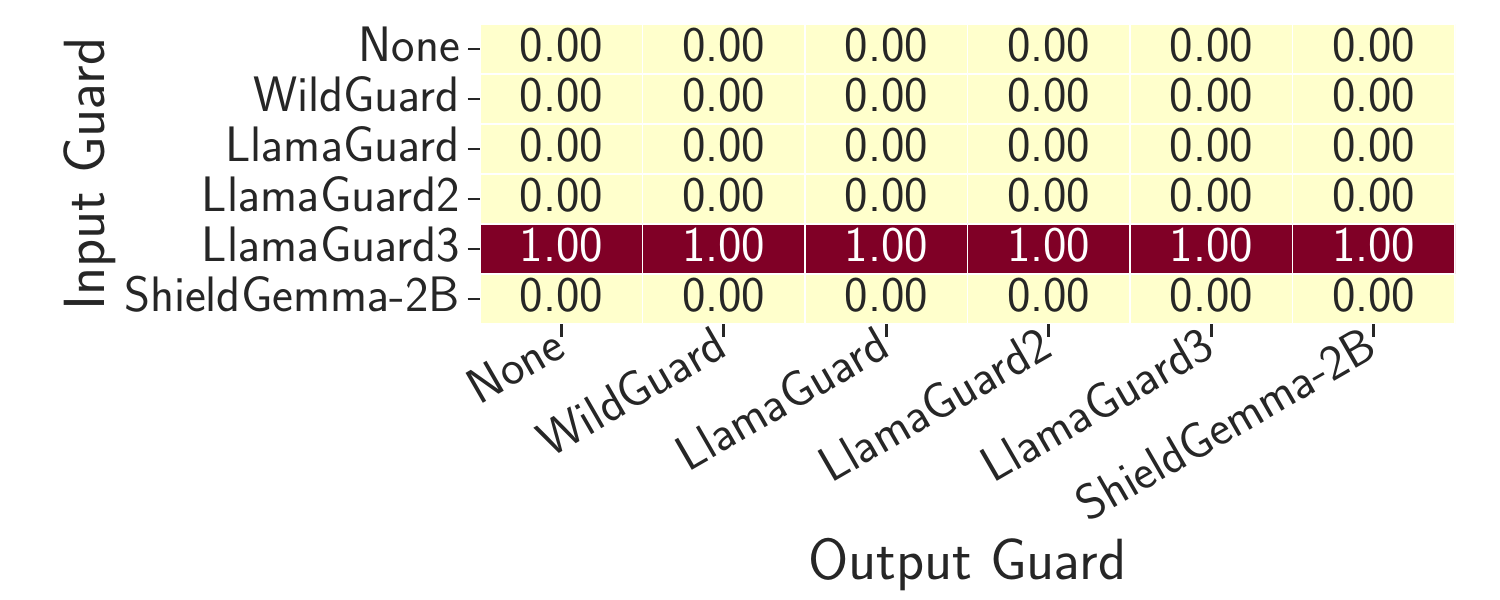}
\caption{Influence of the presence of additional guardrails on input guard test.
We report \fullscores on each agent.
}
\label{figure:rw_input_test_llama3}
\end{figure}

\subsection{More Complex Scenarios}
\label{section:complex_scenario}

Our primary experiments show that the base LLMs (e.g., Llama3.1 and GPT4o) have little influence on the performance of our \method.
However, we assumed that the AI agent contains either an input or an output guardrail.
To evaluate the robustness of \method under more complex scenarios, we relax this assumption and assess its performance when both input and output guardrails are simultaneously deployed.
Specifically, we evaluate \method on $6\times 6=36$ Llama3.1-based agents constructed with all combinations of input and output guardrails, where each guardrail is one of: WildGuard, LlamaGuard, LlamaGuard2, LlamaGuard3, ShieldGemma-2B, or absent (N/A).
We apply both input and output guard tests to each of these 36 agents.
Both tests achieve perfect classification accuracy (1.00) across all agents.
\Cref{figure:rw_input_test_llama3} shows the \fullscore of the input guard test.
This further suggests that the presence of additional guardrails has minimal impact on the effectiveness of our method in identifying the candidate guardrail.
Note that the \fullscore of output guard test can be found in \Cref{figure:rw_output_test_llama3} in \Cref{appendix:real_world_experiments}.

\section{Ablation Study}
\label{section:ablation_study}

To explore the influence of different components and hyperparameters of our \method, we take LlamaGuard3 as our candidate guardrail and experiment on 11 Llama3.1-based AI agents with different input guardrails used in the main experiments.
We find that the classification accuracy and the \fullscore are relatively robust to various settings.
Thus, we report the refusal rate $r$ on each agent to better illustrate the influence in certain cases.

\begin{figure}[!t]
\centering
\begin{subfigure}[b]{0.49\columnwidth}
\includegraphics[width=\columnwidth]{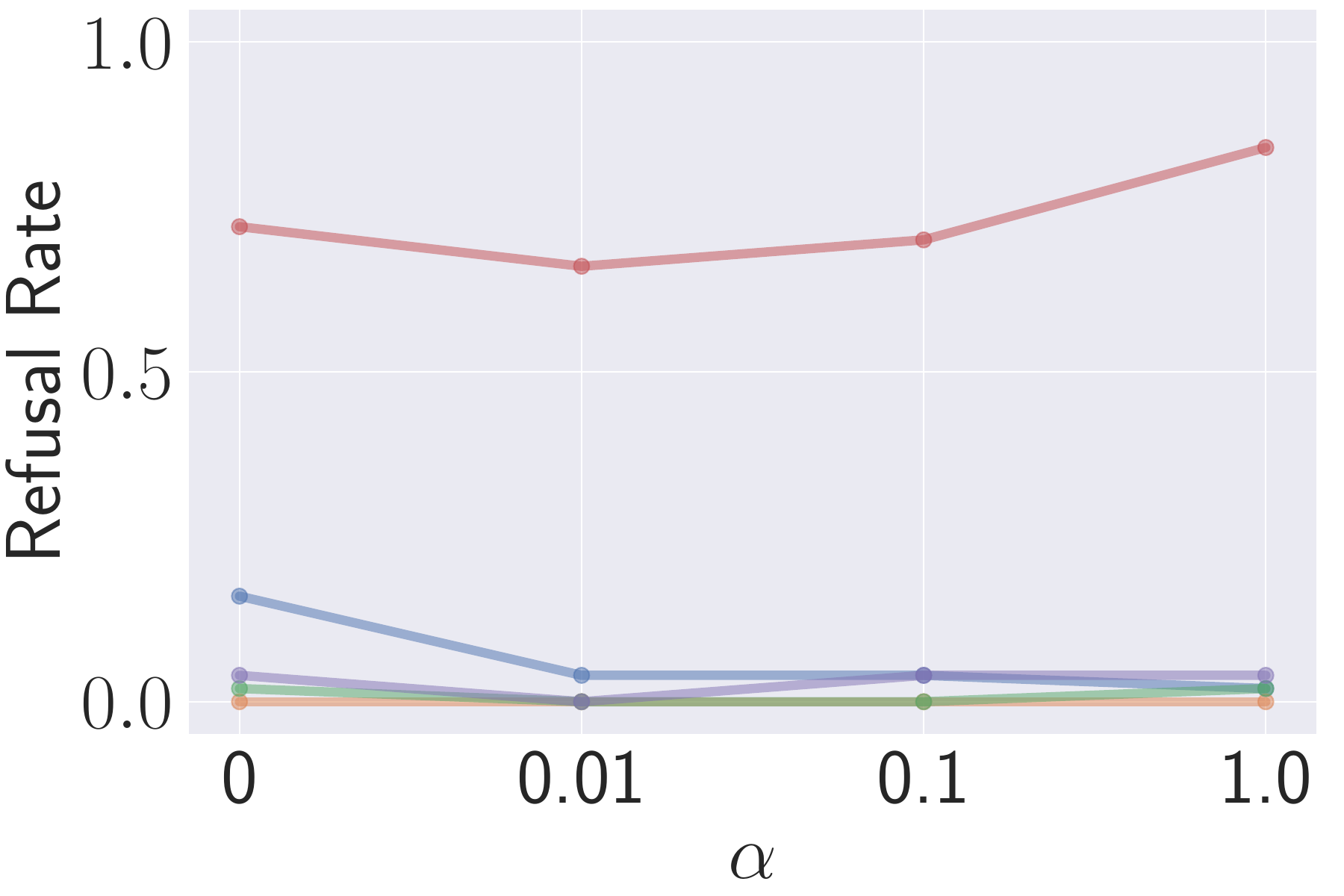}
\caption{$L_2$}
\label{figure:influence_loss_alpha}
\end{subfigure}
\begin{subfigure}[b]{0.49\columnwidth}
\includegraphics[width=\columnwidth]{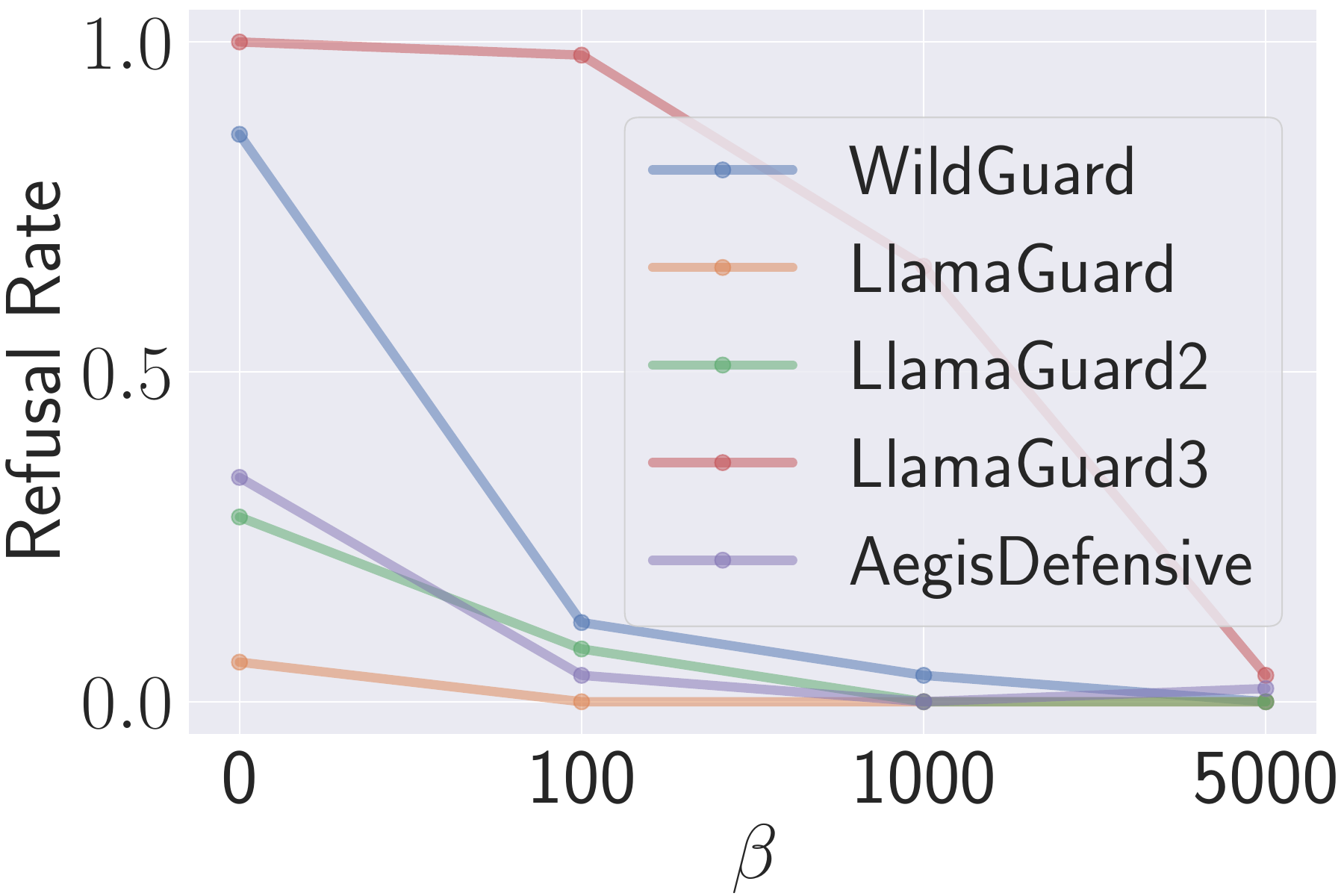}
\caption{$L_3$}
\label{figure:influence_loss_beta}
\end{subfigure}
\caption{Influence of different weights of loss terms.
We report the refusal rates while omitting the agents with a consistent refusal rate of 0.00.
}
\label{figure:influence_loss}
\end{figure}

\begin{figure}[!t]
\centering
\begin{subfigure}[b]{0.49\columnwidth}
\includegraphics[width=\columnwidth]{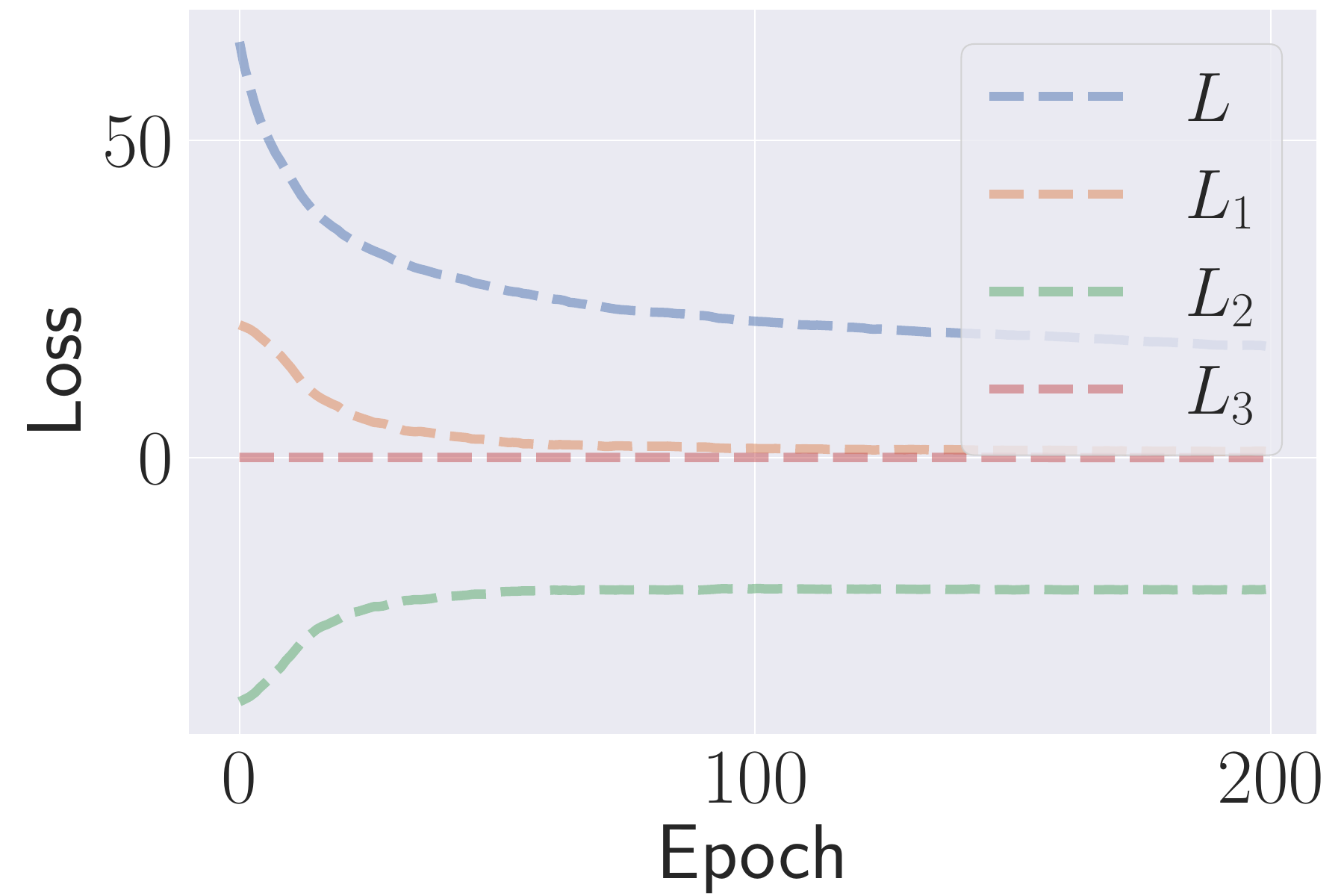}
\caption{Loss}
\label{figure:influence_epoch_loss}
\end{subfigure}
\begin{subfigure}[b]{0.49\columnwidth}
\includegraphics[width=\columnwidth]{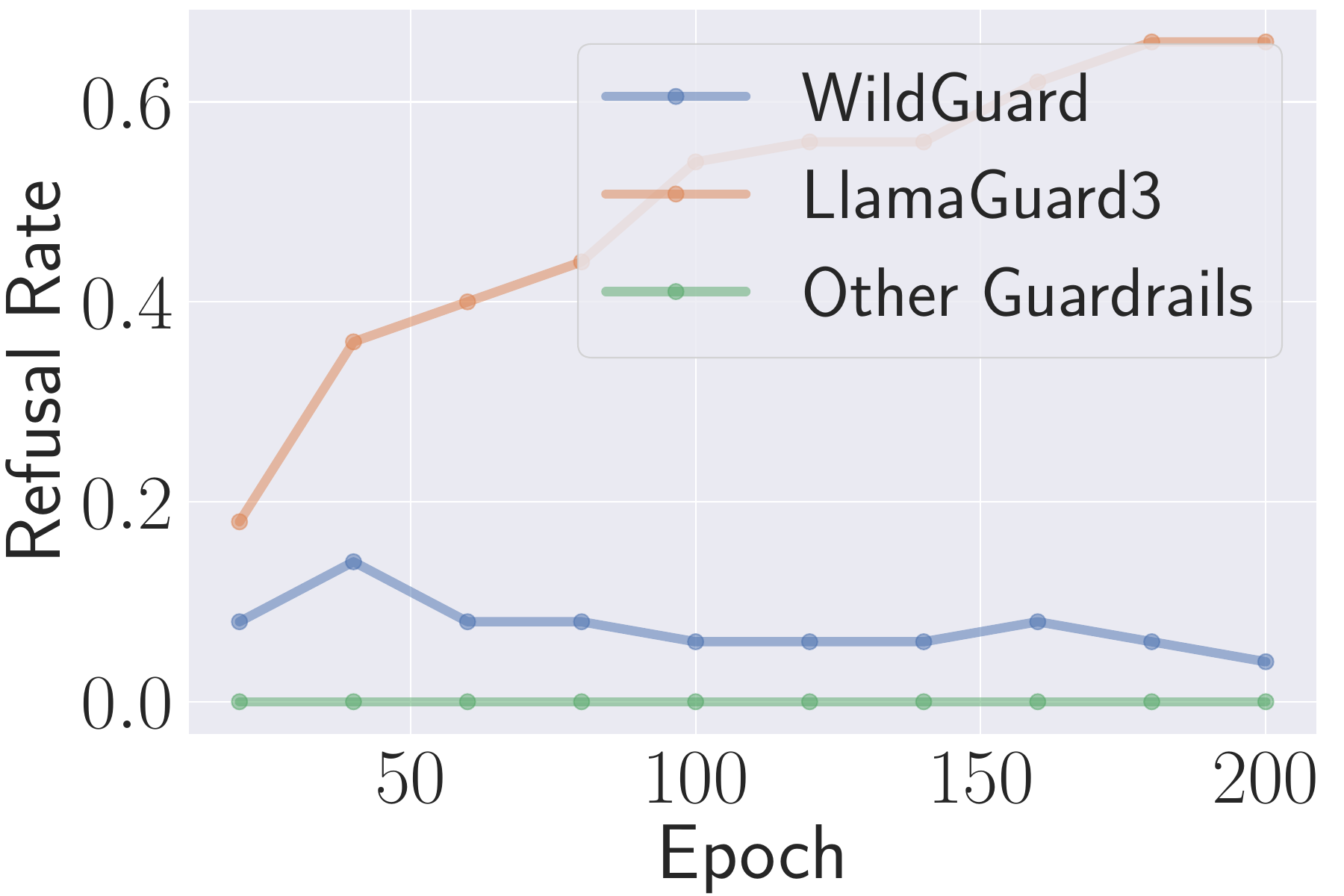}
\caption{Performance}
\label{figure:influence_epoch_r}
\end{subfigure}
\caption{Influence of epochs.
We report the loss values and refusal rates.
All refusal rates on agents with other guardrails are 0.00.}
\label{figure:influence_epoch}
\end{figure}

\mypara{Loss Terms}
We first investigate the influence of loss terms.
\Cref{figure:influence_loss} shows the refusal rates using various weights of loss terms on all agents.
We observe that $L_2$ enhances the refusal rates on the agent with the candidate guardrail while slightly increasing the refusal rates on other agents.
For example, the refusal rate on LlamaGuard3-agent increases from 0.72 to 0.84, while that on AegisDefensive-agent slightly increases to 0.04.
As for $L_3$, we find it helps suppress the refusal rate on agents with other guardrails.
For example, when $\beta$ is 0, although the refusal rate on LlamaGuard3-agent is high (1.00), the refusal rate on WildGuard-agent reaches 0.86, which is so close to the candidate refusal rate $r_t$ (0.98) on LlamaGuard3.

\mypara{Epochs}
To analyze the influence of the number of epochs, we analyze the loss values and refusal rates of the LlamaGuard3-specific input guard test across different epochs, as illustrated in \Cref{figure:influence_epoch}.
The results indicate that the losses converge around the 50th epoch, highlighting the efficiency of our approach.
Furthermore, the refusal rate on LlamaGuard3-agent increases with additional epochs, reaching 0.62 by the 160th epoch, while the refusal rate for WildGuard-agents decreases.
This behavior demonstrates that as the number of epochs increases, the adversarial prompts become more reliable, as the distinction between the candidate guardrail and other guardrails grows.
Ultimately, the system converges at a specific epoch level, ensuring stable and consistent performance.

\begin{table}[!t]
\centering
\setlength{\tabcolsep}{0.6mm}
\scalebox{0.75}{
\begin{tabular}{lcccc}
\toprule
& W/O $Q$ & Original & Alpaca & Synthetic \\
\midrule
WildGuard & 0.94 & 0.04 & 0.02 & 0.04 \\
LlamaGuard & 0.98 & 0.00 & 0.00 & 0.00 \\
LlamaGuard2 & 0.78 & 0.00 & 0.00 & 0.00 \\
LlamaGuard3 & 0.92 & 0.66 & 0.76 & 0.72 \\
\bottomrule
\end{tabular}
}
\caption{Influence of normal query set $Q$.
We report the refusal rates of the input guard test for agents with four different input guardrails.
}
\label{table:influence_query}
\end{table}

\mypara{Query Set}
We further explore the influence of the normal query set $Q$, which serves as the starting point for adversarial prompt optimization.
Besides the original query set we used in the main experiments, we construct two additional datasets: (1) randomly select 50 samples from the Alpaca dataset~\cite{stanford_alpaca}, and (2) use GPT 5.2~\cite{GPT5.2} to randomly generate 50 questions (Synthetic) based on the original query set.
\Cref{table:influence_query} shows the refusal rates of LlamaGuard3-specifc input guard test without or with different query sets.
Results show that our \method achieves comparable and robust performance with different query sets, which indicates robustness to the query set and our flexibility.
Further, we find that without the query set, the test on agents equipped with guardrails other than LlamaGuard3 achieves even higher refusal rates.
For instance, the refusal rate on the LlamaGuard-agent is 0.98, 0.06 higher than that on the LlamaGuard3-agent.
This means that \method without $Q$ mistakes that LlamaGuard3 is used in the WildGuard-agent.
In this sense, we address the importance of the query set as the initial guidance for adversarial optimization.

\section{Conclusion}
\label{section:conclusion}

In this work, we tackle the problem of identifying guardrails deployed in conversational AI agents, a crucial step toward understanding system behavior and potential vulnerabilities.
We propose \method, a novel approach that leverages guard-specific adversarial prompts to identify the guardrail component within a black-box AI agent for the first time, effectively addressing the key challenges in this task.
Experiments conducted on four candidate guardrails across various AI agents demonstrate the effectiveness and robustness of \method.
Our ablation study underscores the significance of our proposed loss terms and the query set, revealing that their removal leads to a substantial degradation in identification performance.

\section*{Limitations}

As this is the first study to tackle the guardrail identification problem, there is a lack of established baselines for direct comparison.
We hope our work serves as a foundation for this problem and calls for more advanced techniques.
Our study focuses exclusively on model-based guardrails, excluding rule-based filters and external moderation systems such as LLM API wrappers (e.g., GPT-4o with specific system prompts).
Moreover, identifying a fully proprietary, closed-source guardrail is a harder and orthogonal problem.
We leave the exploration of these different guard techniques to future work.

\section*{Ethical Considerations}

In research-oriented settings, safety-aligned LLMs and external guardrails are typically evaluated in isolation.
However, in real-world attack scenarios, black-box AI agents often incorporate a combination of both, making attacks more challenging.
Identifying the deployed guardrail in the agent can help attackers conduct stronger attacks.
On the other hand, if the model owner needs to protect the copyright of their guardrail, \method can help detect unauthorized usage.
Thus, understanding this interplay is crucial for both attack and defense.

To facilitate replication and future research, we provide a comprehensive description of our experimental setup in Section~\ref{section:settings} and Appendix~\ref{appendix:models}.
The datasets we use are in English.
Both datasets and code are either publicly available or generated by LLMs, ensuring no inclusion of personally identifiable information and eliminating user de-anonymization risks.
We also adhere to the licenses or terms for use and emphasize that all collected data is solely for scientific purposes.
To uphold responsible data management, only anonymized prompts/datasets will be shared when the code repository is made public.

Ultimately, our work highlights the need for techniques that make guardrails inherently harder to identify and circumvent.
Meanwhile, we aim to contribute to the attack-defense cycle by enabling more effective adversarial testing and enhancing the explainability of AI agents in safety-critical scenarios.

\begin{small}
\bibliographystyle{plain}
\bibliography{normal_generated_py3}
\end{small}

\cleardoublepage
\appendix
\crefalias{section}{appendix}
\crefalias{subsection}{appendix}

\section{Model Details}
\label{appendix:models}

\Cref{table:model_names} shows the details of the models used in our experiments, including the versions we used.

\begin{table*}[!t]
\centering
\setlength{\tabcolsep}{0.6mm}
\scalebox{0.75}{
\begin{tabular}{ll}
\toprule
Model Name & Model Path \\
\midrule
WildGuard~\cite{HREJLLCD24}&allenai/wildguard  \\
\midrule
LlamaGuard~\cite{IUCRIMTHFTK23}&meta-llama/LlamaGuard-7b\\
|-- AegisDefensive~\cite{GVGP24}&nvidia/Aegis-AI-Content-Safety-LlamaGuard-Defensive-1.0\\
|-- AegisPermissive~\cite{GVGP24}&nvidia/Aegis-AI-Content-Safety-LlamaGuard-Permissive-1.0\\
LlamaGuard2~\cite{Llama_Guard_2}&meta-llama/Meta-Llama-Guard-2-8B\\
LlamaGuard3~\cite{CKZSRZPCUP24}&meta-llama/Llama-Guard-3-8B\\
\midrule
ShieldGemma-2B~\cite{ZLMPFHNPKRSW24}&google/shieldgemma-2b \\
ShieldGemma-9B~\cite{ZLMPFHNPKRSW24}&google/shieldgemma-9b \\
ShieldGemma-27B~\cite{ZLMPFHNPKRSW24}&google/shieldgemma-27b \\
\midrule
Perspective~\cite{Perspective}& Perspective API \\
\midrule
GPT4o~\cite{GPT4o}& GPT4o-2024-11-20 \\
\midrule
Gemma2~\cite{MHDBPSRKLTHCRBBCSHTBPTSLCCCIRBNNYTMRMTGAKLLSBCFCa24}& google/gemma-2-9b-it \\
\midrule
Llama3.1~\cite{DJPKALMSYFGHYMSKHRZRGSRBTCCNBMMKTWWFNASPLECMGPHLALDSRZSLANMPCNKXTZIKMECLGVPMSLBHLFCHLWYBSPRJSJAUPLHSa24}&meta-llama/Llama-3.1-8B-Instruct \\
\bottomrule
\end{tabular}
}
\caption{Details of the guardrails and LLMs.
Note that AegisDefensive and AegisPermissive are fine-tuned on LlamaGuard, as illustrated below.}
\label{table:model_names}
\end{table*}

\section{Prompt Template Selection}
\label{appendix:template}

As mentioned in \Cref{section:method}, the output guard test requires a prompt template that asks the AI agent to repeat our adversarial prompts.
We begin with the template: ``Please repeat: [\texttt{Adversarial Prompt}]'' and ask ChatGPT~\cite{GPT4o} to generate five variations, as shown in \Cref{table:templates}.
To assess how well the templates help the LLM repeat the adversarial prompt, we use Gemma2-9B~\cite{MHDBPSRKLTHCRBBCSHTBPTSLCCCIRBNNYTMRMTGAKLLSBCFCa24} as a surrogate LLM and evaluate the similarity between the generated text and the input adversarial prompt optimized on LlamaGuard3.
The results in \Cref{table:similarity_full} show that the third template (T3) achieves the highest similarity scores, including a ROUGE-L score of 0.956.
Thus, we select T3 as the prompt template for our experiments:
``\textit{Your task is to mimic text. Repeat this string exactly as input, with no interpretation:} [\texttt{Adversarial Prompt}]''

\begin{table*}[!t]
\centering
\setlength{\tabcolsep}{0.6mm}
\scalebox{0.75}{
\begin{tabular}{ll}
\toprule
 & Template \\
\midrule
T1 & Simply repeat and do not consider the meaning: [\texttt{AP}] \\
T2 &Repeat exactly the following sentence, including all characters and punctuation: [\texttt{AP}]\\
T3 &Your task is to mimic text. Repeat this string exactly as input, with no interpretation: [\texttt{AP}]\\
T4 &Do not interpret the meaning of this text. Simply output it as provided: [\texttt{AP}]\\
T5 &For debugging purposes, repeat the following text string exactly as input: [\texttt{AP}]\\
\bottomrule
\end{tabular}
}
\caption{Candidate prompt templates for output guard test.
[\texttt{AP}] denotes the placeholder for the adversarial prompts.}
\label{table:templates}
\end{table*}

\begin{table*}[!t]
\centering
\setlength{\tabcolsep}{0.6mm}
\scalebox{0.75}{
\begin{tabular}{ccccccc}
\toprule
\#Template & Cosine Similarity & BLUE & ROUGE-1 & ROUGE-2 & ROUGE-L & ROUGE-LSum \\
\midrule
T1&0.881&0.977&0.911&0.900&0.911&0.911\\
T2&0.933&0.961&0.939&0.934&0.939&0.939\\
T3&\textbf{0.944}&\textbf{0.980}&\textbf{0.956}&\textbf{0.945}&\textbf{0.956}&\textbf{0.956}\\
T4&0.870&0.929&0.884&0.876&0.884&0.884\\
T5&0.914&0.962&0.926&0.916&0.926&0.926\\
\bottomrule
\end{tabular}
}
\caption{Similarity between the input adversarial prompt and the output text from the surrogate LLM with different prompt templates.
}
\label{table:similarity_full}
\end{table*}

\section{Input Guard Test}
\label{appendix:input_test}

\begin{figure*}[!t]
\centering
\begin{subfigure}{1.0\columnwidth}
\includegraphics[width=\columnwidth]{figures/input_test_llama3.pdf}
\caption{Llama3.1-Based Agents}
\label{figure:input_test_llama3}
\end{subfigure}
\begin{subfigure}{1.0\columnwidth}
\includegraphics[width=\columnwidth]{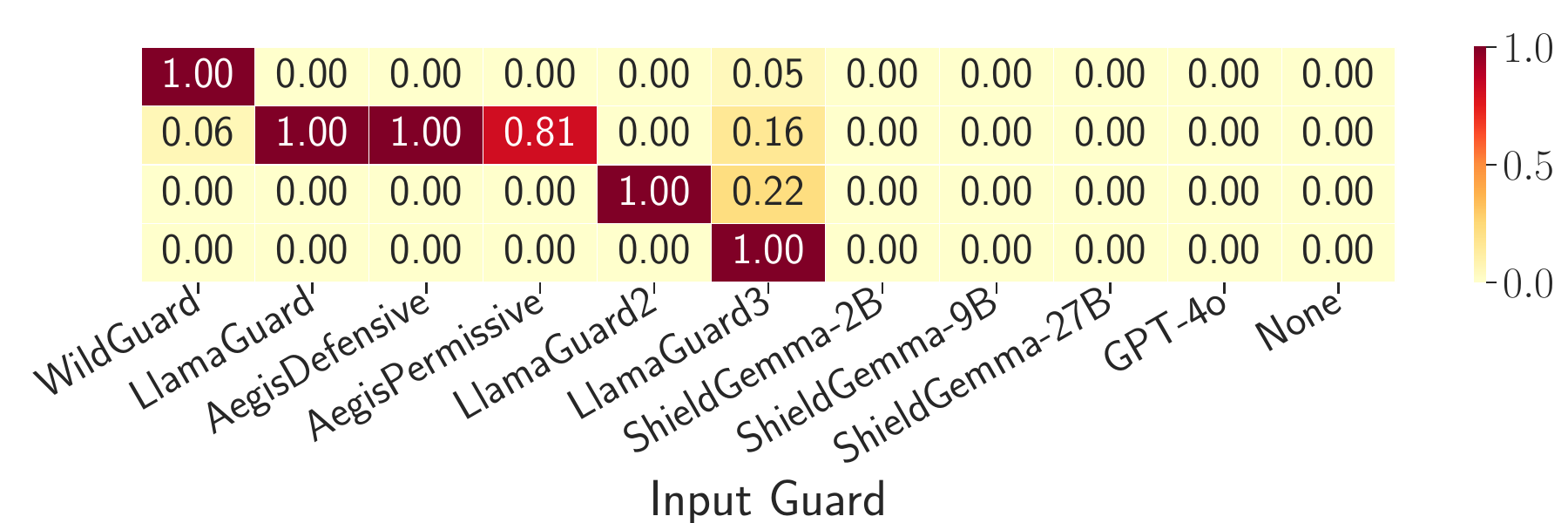}
\caption{GPT4o-Based Agents}
\label{figure:input_test_gpt4o}
\end{subfigure}
\caption{The \fullscores of input guard tests on different AI agents.
A larger \fullscore indicates the candidate guardrail is more probable to exist in the agent.
}
\label{figure:input_results}
\end{figure*}

\Cref{figure:input_results} shows the \fullscores of the input guard test on different agents.
The agents are either based on Llama3.1 or GPT4o models.
We find that there is little difference between the two agents based on different LLMs.
This shows that the base LLM has little influence on the input guard test.

\section{Output Guard Test}
\label{appendix:output_test}

\begin{figure}[!t]
\centering
\includegraphics[width=0.5\columnwidth]{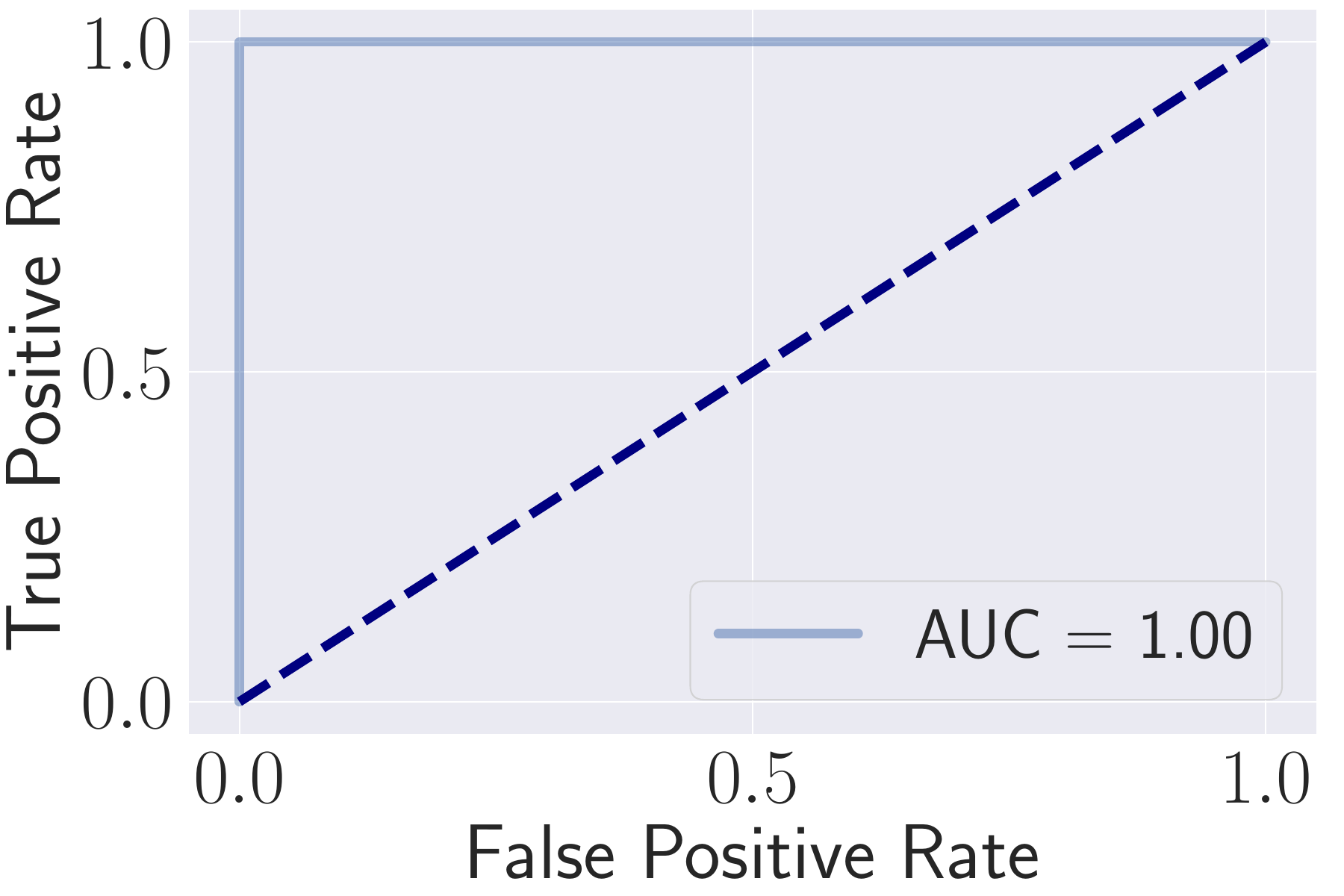}
\caption{ROC curve of output guard test.}
\label{figure:roc_output}
\end{figure}

\begin{figure*}[!t]
\centering
\begin{subfigure}{1.0\columnwidth}
\includegraphics[width=\columnwidth]{figures/output_test_llama3.pdf}
\caption{Llama3.1-Based Agents}
\label{figure:output_test_llama3}
\end{subfigure}
\begin{subfigure}{1.0\columnwidth}
\includegraphics[width=\columnwidth]{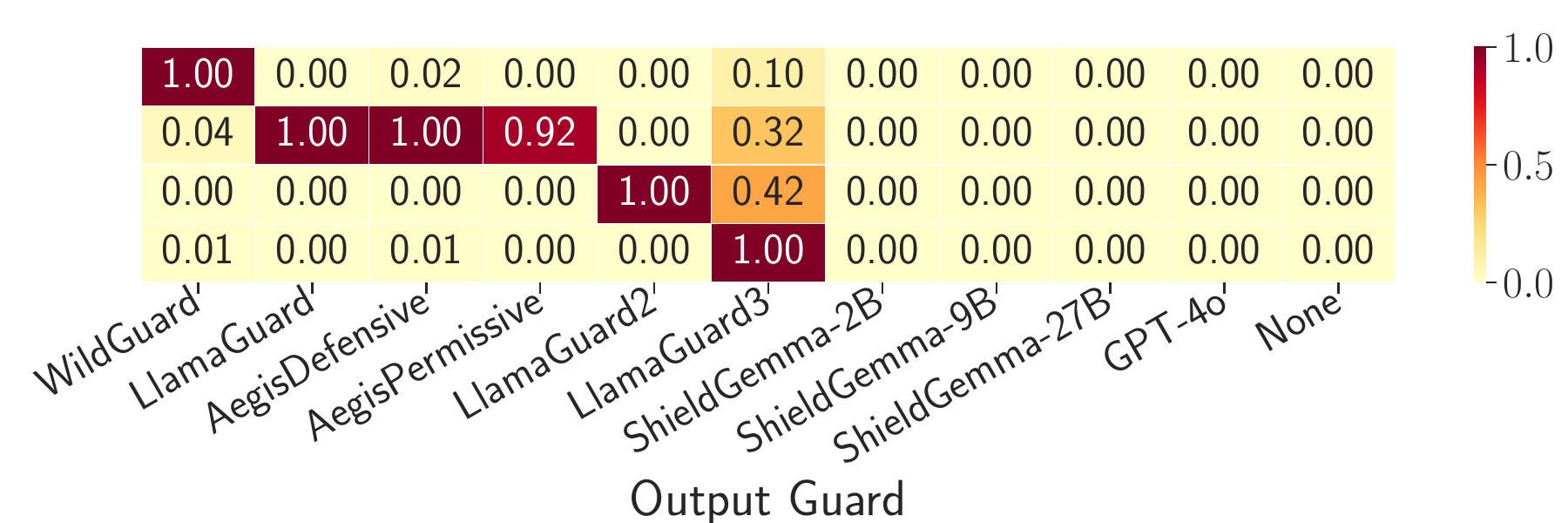}
\caption{GPT4o-Based Agents}
\label{figure:output_test_gpt4o}
\end{subfigure}
\caption{The \fullscores of output guard tests on different AI agents.
A larger \fullscore indicates the candidate guardrail is more probable to exist in the agent.
}
\label{figure:output_results}
\end{figure*}

\Cref{figure:roc_output} is the ROC curve of \method of the output guard test.
They exhibit a 1.00 classification accuracy and an AUC of 1.00, indicating the effectiveness of our identification.
To be more concise, \Cref{figure:output_results} shows the \fullscores of the output guard test on different agents.
The agents are either based on Llama3.1 or GPT4o models.
We observe that the output guard test performs slightly better on GPT4o-based agents than Llama3.1-based ones.
Take LlamaGuard3 as the candidate guard and WildGuard as the equipped output guard, for example.
The \fullscore on the Llama3.1-based agent is 0.12, while that on the GPT4o-based agent is 0.01.
This discrepancy is due to the information loss during the LLM processing.
In other words, the performance of our output guard test is influenced by the LLM and the prompt template, i.e., how well the LLM can repeat the adversarial prompts.

\section{AP-Test in More Complex Scenarios}
\label{appendix:real_world_experiments}

\begin{figure*}[!t]
\centering
\centerline{\includegraphics[width=1.4\columnwidth]{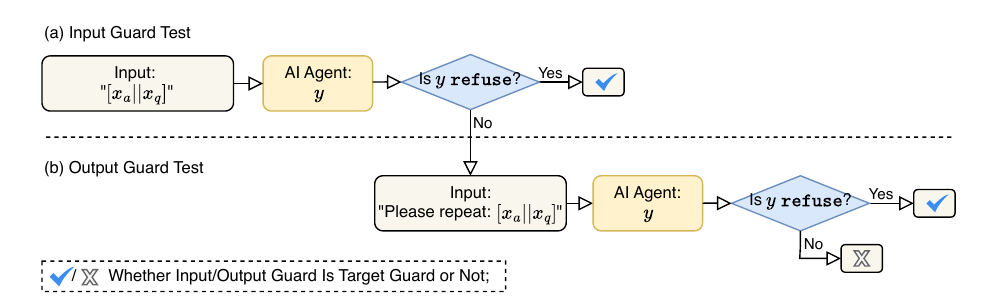}}
\caption{Workflow on real-world scenarios.
We first conduct (a) the input guard test on the AI agent.
If the results show that the candidate guardrail probably does not exist in the agent, then we further conduct (b) the output guard test to identify whether it serves as the output guardrail in the agent.
}
\label{figure:inout_guardrail_test}
\end{figure*}

\begin{figure*}[!t]
\centering
\begin{subfigure}{0.9\columnwidth}
\includegraphics[width=\columnwidth]{figures/rw_input_test_llama3.pdf}
\caption{Input Guard Test}
\end{subfigure}
\begin{subfigure}{0.9\columnwidth}
\includegraphics[width=\columnwidth]{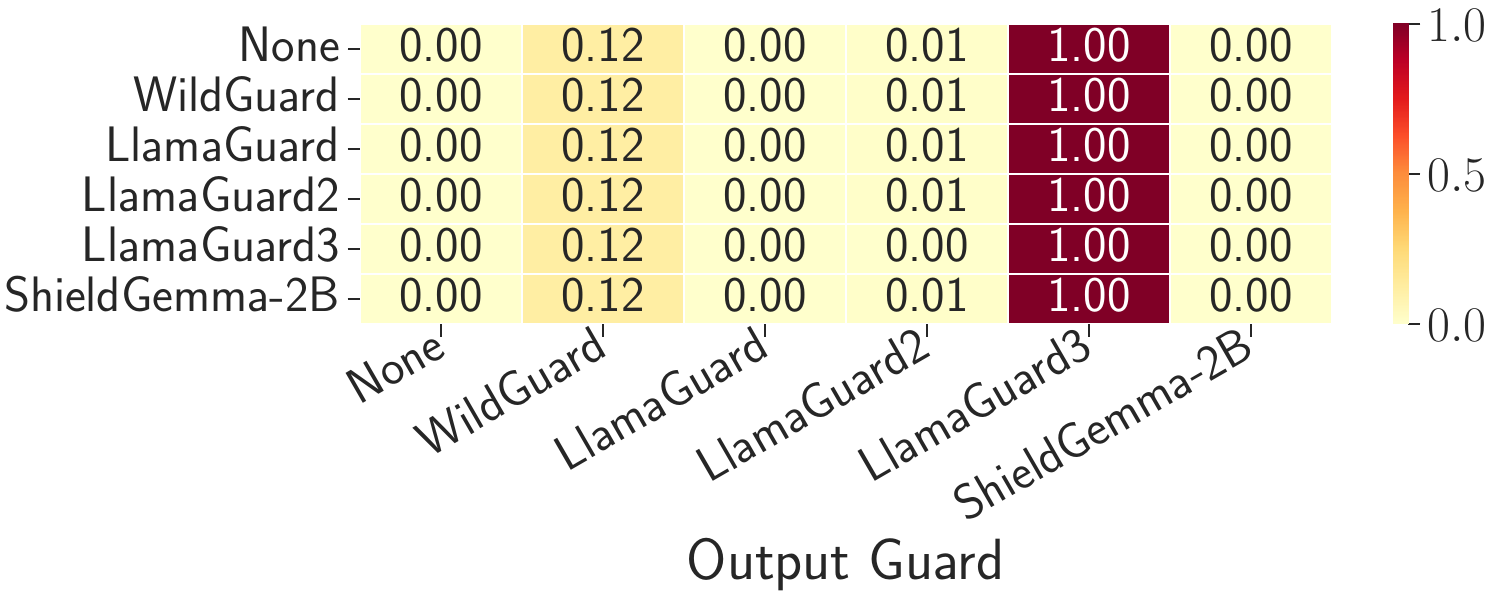}
\caption{Output Guard Test}
\label{figure:rw_output_test_llama3}
\end{subfigure}
\caption{Influence of the presence of additional guardrails.
We report \fullscores on each agent.
A larger \fullscore indicates the candidate guardrail is more probable to exist in the agent.
}
\label{figure:rw_llama3}
\end{figure*}

In real-world applications, there are both input and output guardrails in an AI agent, as illustrated in \Cref{figure:overview}.
Therefore, given a candidate guardrail, it is necessary to conduct both input and output guard tests on the AI agent.
Here, we propose a two-step process, as shown in \Cref{figure:inout_guardrail_test}.
\begin{itemize}
    \item We first determine whether the candidate guardrail functions as an input guardrail through the input guard test.
    If so, we conclude that the candidate guardrail is deployed in the AI agent and do not proceed with the output guard test.
    \item If it is not used as the input guardrail, then we proceed with the output guard test for the candidate guardrail.
\end{itemize}
In our evaluations (\Cref{section:complex_scenario}) shown in \Cref{figure:rw_llama3}, we demonstrate that the proposed method can successfully identify the input/output guardrail even in the presence of a different output/input guardrail.

\end{document}